\documentclass[a4paper,11pt]{article}
\usepackage{mathrsfs}
\usepackage{amssymb}
\usepackage{stmaryrd}
\usepackage{txfonts}
\usepackage[a4paper,left=3cm,right=3cm]{geometry}
\usepackage{graphicx}

\newcommand{\be}{\begin{equation}}
\newcommand{\ee}{\end{equation}}
\newcommand{\bea}{\begin{eqnarray}}
\newcommand{\eea}{\end{eqnarray}}

\begin{document}
\setlength{\unitlength}{1.0mm}
\title{Gauge Invariance of the Muonium-Antimuonium Oscillation Time Scale and Limits on Right-Handed Neutrino Masses}
\author{Boyang Liu\thanks{liu115@physics.purdue.edu}\\
\small \emph{Department of Physics, Purdue University, West
Lafayette, IN 47906,USA}}
\date{}
\maketitle \normalsize
\begin{abstract}
The gauge invariance of the muonium-antimuonium ($M\bar{M}$)
oscillation time scale is explicitly demonstrated in the Standard
Model modified only by the inclusion of singlet right-handed
neutrinos and allowing for general renormalizable interactions. The
see-saw mechanism is exploited resulting in three light Majorana
neutrinos and three heavy Majorana neutrinos with mass scale $M_R\gg
M_W$. The leading order matrix element contribution to the
$M\bar{M}$ oscillation process is computed in $R_\xi$ gauge and
shown to be $\xi$ independent thereby establishing the gauge
invariance to this order. Present experimental limits resulting from
the non-observation of the oscillation process sets a lower limit on
$M_R$ roughly of order $600$ GeV.
\end{abstract}

\section{Introduction}
Muonium $(M)$ is the Coulombic bound state of an electron and an
antimuon $(e^-\mu^+)$, while antimuonium $(\bar M)$ is the Coulombic
bound state of a positron and a muon $(e^+\mu^-)$. It was suggested
roughly 50 years ago\cite{Pontecorvo} that there may be a
spontaneous conversion between muonium and antimuonium which would
violate the individul electron and muon number conservation laws by
two units. Such a muonium-antimuonium oscillation is totally
forbidden within the Standard Model. Hence, its observation will be
a clear signal of physics beyond the Standard Model. Since the
initial suggestion, experimental searches have been
conducted\cite{Huber}-\cite{Willmann} and a variety of theoretical
models have been proposed which could give rise to such a
muonium-antimuonium conversion. These include interactions which
could be mediated by (a) a doubly charged Higgs boson
$\Delta^{++}$\cite{Halprin,Herczeg}, which is contained in a
left-right symmetric model, (b) massive Majorana
neutrinos\cite{Clark&Love, Kim}, or (c) the $\tau$-sneutrino in an
R-parity violation supersymmetric model\cite{Halprin and Masiero}.

In this paper, we focus on  a modified Standard Model which includes
singlet right-handed neutrinos. There is now compelling evidence of
the existence of neutrino oscillations from the experimental study
of atmospheric and solar neutrinos\cite{Fukuda}-\cite{Ahn}. That
implies nonzero neutrino masses and mixing matrix elements. The size
and nature of the neutrino mass and the associated mixing is still
an open question subject to experimental determination and
theoretical speculation \cite{Altarelli}-\cite{Gonzalez}. One simple
neutrino mass model is obtained by modifying the Standard Model by
including singlet right-handed neutrinos and allowing for a general
mass matrix for neutrinos. Left-handed neutrinos along with their
charged leptonic partners are components of $SU(2)_L$ doublets and
experience the weak interaction while any right-handed neutrinos are
completely neutral under the Standard Model gauge group. The see-saw
mechanism\cite{Minkowski}-\cite{Mohapatra} provides a natural
explanation of the smallness of the three light Majorana neutrino
masses, while ensuring that the other three Majorana neutrinos are
heavy. Such a model could also lead to the muonium-antimuonium
oscillation process. In order for there to be a nontrivial mixing
between muonium and antimuonium, the individual electron and muon
number conservation must be violated. Such a situation results
provided the neutrinos are massive particles which mix amongst the
various generations. This criterion can be met by the modified
Standard Model and the $e^-\mu^+$ and $e^+\mu^-$ states could indeed
mix.

\section{Neutrino masses and mixings}
The leptonic Yukawa interactions with the Higgs scalar doublet in the modified Standard Model take the form
\be {\cal L}^\phi_{int}=-\frac{g}{\sqrt 2
M_W}\phi^-\Big(\sum^3_{a,b=1}\overline{\ell^{(0)}_{Ra}}
m^\ell_{ab}\nu^{(0)}_{Lb}-\overline{\ell^{(0)}_{La}}{m^D_{ab}}^\dagger\nu^{(0)}_{Rb}\Big)+H.C.\ee
Here $\ell^{(0)}_{La}$ and $\nu^{(0)}_{La}$ are respectively the
charged lepton and its associated neutrino partner of the $SU(2)_L$
doublet, while $\nu^{(0)}_{Ra}$ is the right-handed neutrino
singlet. The superscript zero indicates weak interaction
eigenstates so that the leptonic charged current interaction is
\be {\cal {L}}^W_{int}
 =-\frac{g}{\sqrt 2}W^{-\mu}\sum^3_{a=1}\overline{\ell^{(0)}_{La}}\gamma_\mu\nu^{(0)}_{La}+H.C.\ee

After spontaneous symmetry
breaking, the mass term for the charged leptons takes the form\be {\cal
L}^\ell_{mass}=-\sum^3_{a,b=1}
[\overline{\ell^{(0)}_{Ra}}m^\ell_{ab}\ell^{(0)}_{Lb}+\overline{\ell^{(0)}_{La}}m^{\ell\ast}_{ba}\ell^{(0)}_{Rb}]\ee
where $m^\ell$ is a $3\times3$ mass matrix. To diagonalize this
matrix, one performs the biunitary transformation \be
m^\ell=A^Rm^\ell_{diag}(A^L)^\dagger\ee where $A^R$ and $A^L$ are
$3\times3$ unitary matrices and $m^\ell_{diag}$ is a diagonal matrix
whose entries are the charged lepton masses. To implement this basis
change, the charged lepton fields participating in the weak
interaction are rewritten in terms of the mass diagonal fields as\be
\ell^{(0)}_{La}=\sum^3_{b=1}A^L_{ab}\ell_{Lb},~~~~~~~~~\ell^{(0)}_{Ra}=\sum^3_{b=1}A^R_{ab}\ell_{Rb}\ee
So doing the mass term reads\be  {\cal
L}^\ell_{mass}=-\sum^3_{a=1}m_{\ell a}
[\overline{\ell_{Ra}}\ell_{Lb}+\overline{\ell_{La}}\ell_{Rb}]\ee A general neutrino mass term resulting from renormalizable
interactions takes the form \be {\cal L}^\nu_{mass}=
-\frac{1}{2}\left(\begin{array}{c}\overline{(\nu^{(0)}_L)^c}
~~\overline{\nu^{(0)}_R}\end{array}\right)
\left(\begin{array}{c}0~~~~~~ (m^D)^T\\m^D~~~~~~m^R\end{array}\right)\left(\begin{array}{c}\nu^{(0)}_L\\
(\nu^{(0)}_R)^c\end{array}\right)+H.C.\ee Note that the upper left
$3\times 3$ block in the neutrino mass matrix is set to zero.  This
block matrix involves only left-handed neutrinos and in the (modified) Standard
Model its generation requires a nonrenormalizable mass
dimension-five operator. Consequently such a term will be ignored.

For three generations of neutrinos, the six mass eigenvalues,
$m_{\nu A}$, are obtained from the diagonalization of the $6\times6$
matrix \be M^\nu=\left(\begin{array}{c}0~~~~~~
(m^D)^T\\m^D~~~~~~m^R\end{array}\right)\ee Since $M^\nu$ is
symmetric, it can be diagonalized by a single unitary $6\times 6$
matrix, $U$, as \be M^\nu_{diag}=U^TM^\nu U.\ee This diagonalization
is implemented via the basis change on the original neutrino fields
organized as the 6 dimensional column vector\be
N^{(0)}_L=\left(\begin{array}{c}
\nu^{(0)}_L\\(\nu^{(0)}_R)^c\end{array}\right),~~~~~~~N^{(0)}_R=\left(\begin{array}{c}
(\nu^{(0)}_L)^c\\\nu^{(0)}_R\end{array}\right)\ee to the new
neutrino fields defined as \be N^{(0)}_L=UN_L, ~~~~~N^{(0)}_R=U^\ast
N_R\ee where \be
N_L=\left(\begin{array}{c}\nu_L\\(\nu_R)^c\end{array}\right),
~~~~~~~N_R= \left(\begin{array}{c}(\nu_L)^c\\
\nu_R\end{array}\right).\ee The neutrino mass term then takes the form\be {\cal
L}^\nu_{mass}=-\frac{1}{2}\sum_{A=1}^6m_{\nu
A}[\nu^T_AC\nu_A+\overline{\nu_A}C\overline{\nu_A^T}]=
-\sum_{A=1}^6m_{\nu A}\overline{\nu_A}\nu_A, \ee where $m_{\nu A}$
are the Majorana neutrino masses.

Since a nonzero Majorana mass matrix $m^R$ does not require $SU(2)_L
\times U(1)$ symmetry breaking, it is naturally characterized by a
much larger scale, $M_R$, than the elements of the matrix $m^D$
whose nontrivial values do require $SU(2)_L\times U(1)$ symmetry
breaking and are thus expected to be somewhere of the order of the
charged lepton mass to the W mass. Thus one can take the elements of
$m^D$, characterized by a scale $m_D$, to be much less than $M_R$,
the scale of the elements of $m^R$. One then finds on
diagonalization of the $6 \times 6$ neutrino mass matrix that three
of the eigenvalues are crudely given by\be m_{\nu
a}\sim\frac{m^2_D}{M_R}\ll m_D, ~~~~a=1,2,3,\ee while the other
three eigenvalues are roughly\be m_{\nu i}\sim M_R,~~~~~i=4,5,6. \ee
This constitutes the so called see-saw
mechanism\cite{Minkowski}-\cite{Mohapatra} and provides a natural
explanation of the smallness of the three light neutrino masses.
Moreover, the elements of the mixing matrix are characterized by an
$M_R$ mass dependence\bea &&U_{ab}\sim \mathcal {O}(1),
~~~~a,b=1,2,3\cr &&U_{ij}\sim\mathcal{O}(1), ~~~~i,j=4,5,6\cr
&&U_{ia}\sim U_{ai}\sim\mathcal{O}(\frac{m_D}{M_R}), ~~~~a=1,2,3,
i=4,5,6. \eea Since the charged lepton mixing matrix is independent
of $M_R$, one finds that elements of the mixing matrix appearing in
the charged current have the $M_R$ mass dependence \bea
&&V_{ab}\sim\mathcal O(1),~~~~ a,b=1,2,3\cr &&V_{ai}\sim\mathcal O
(\frac{m_D}{M_R}),~~~~a=1,2,3, ~~i=4,5,6\eea

Inserting the transformations (5) and (11) in the interaction terms
(1) and (2), and taking into account the mass matrix transformations (4)
and (9), one obtains the explicit interactions of charged bosons with
the leptons in their mass diagonal basis as \be
{\cal{L}}^W_{int}=-\frac{g}{\sqrt
2}W^{-\mu}\sum^3_{a=1}\sum^6_{A=1}\bar \ell_{La}\gamma_\mu
V_{aA}\nu_A-\frac{g}{\sqrt
2}W^{+\mu}\sum^3_{a=1}\sum^6_{A=1}\bar\nu_A
V^\ast_{aA}\gamma_\mu\ell_{La}\ee \bea {\cal
L}^{\phi}_{int}=&&-\frac{g}{\sqrt 2
M_W}\phi^-\sum^3_{a=1}\sum^6_{A=1}\bar{\ell}_{a}V_{aA}~\Big{(}m_{la}
\frac{1-\gamma_5}{2}-m_{\nu A}\frac{1+\gamma_5}{2}\Big{)}~\nu_A\cr
&&-\frac{g}{\sqrt 2
M_W}\phi^+\sum^3_{a=1}\sum^6_{A=1}\bar{\nu}_{A}V_{aA}^\ast~\Big{(}m_{la}
\frac{1+\gamma_5}{2}-m_{\nu A}\frac{1-\gamma_5}{2}\Big{)}~\ell_a\eea
where \be V_{aA}=\sum_{b=1}^3(A^{-1}_L)_{ab}U_{bA}\ee

Note that the mixing matrix $V_{aA}$ satisfies the
identities\cite{Pilaftsis}: \be
\sum_{A=1}^6V_{aA}V^\ast_{bA}=\delta_{ab}\ee \be
\sum^6_{A=1}V_{aA}V_{bA}m_{\nu A}=0 \ee Identity (21) stems from the
unitarity of matrices $A_L$ and $U$, while identity (22) is a consequence of the particular
form of the neutrino mass matrix. In particular, it requires  the
Majorana mass term of the left-handed neutrinos be set to zero. A detailed proof of this later identity is provided in the Appendix.

\section{The gauge invariant T-matrix elements}
The lowest order Feynman diagrams accounting for muonium and antimuonium
mixing are displayed in Fig.1. We shall consistently employ the $R_\xi$ gauge. The  gauge invariance of the T-matrix element
will be demonstrated by establishing its $\xi$ independence. In Fig.1, there are two neutrinos in
the intermediate state for each graph while every wavy line represents
either a W boson or an $R_\xi$ gauge charged erstwhile Nambu-Goldstone boson.
\newline
\begin{center}
\includegraphics[width=14cm]{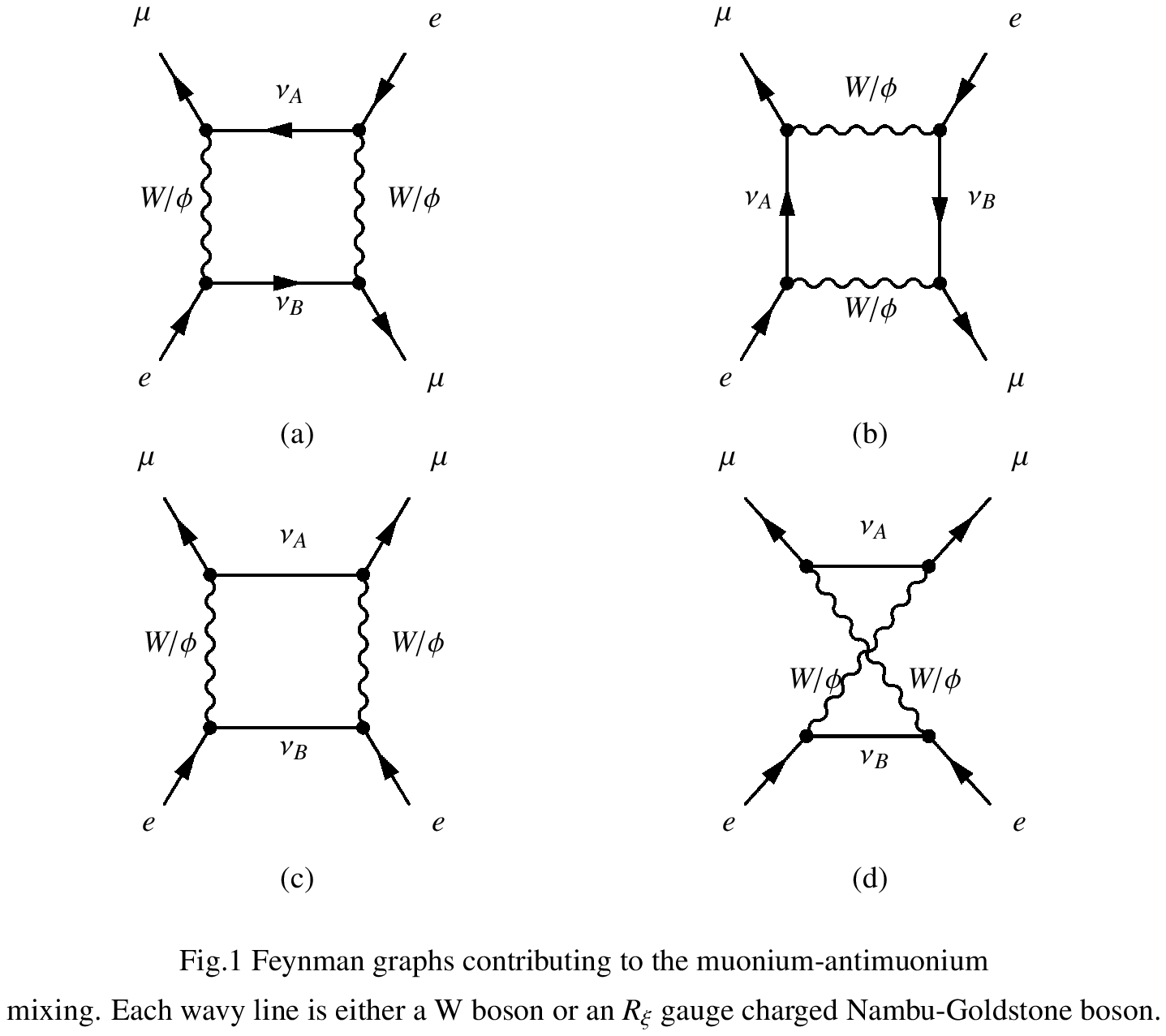}
\end{center}


Note that in unitary gauge ( $\xi\rightarrow
\infty$), the W boson propagator takes
the form $\frac{-i}{p^2-M_W^2+i\epsilon}[g_{\mu\nu}-\frac{p_\mu
p_\nu}{M_W^2}]$. A theory with such a propagator has very bad power counting
convergence properties. As it turns out, the unitary gauge power counting divergent pieces in the $W$ vector box diagrams vanish, as they must,  after application of properties (21) and (22).
Hence, when we calculate the T-matrix elements in $R_\xi$ gauge, we
will also apply properties (21) and (22) to establish the cancellation of the various terms in this case.

As it turns out, graph (a) gives the same contribution as (b), as do
graphs (c) and (d). Hence, we need only
discuss the gauge invariant T-matrix elements of the graphs (a) and
(c). Fig. 2 details explicitly the 4 separate graphs which are represented by the single graph in Fig. 1.
\newline
\begin{center}
\includegraphics[width=12cm]{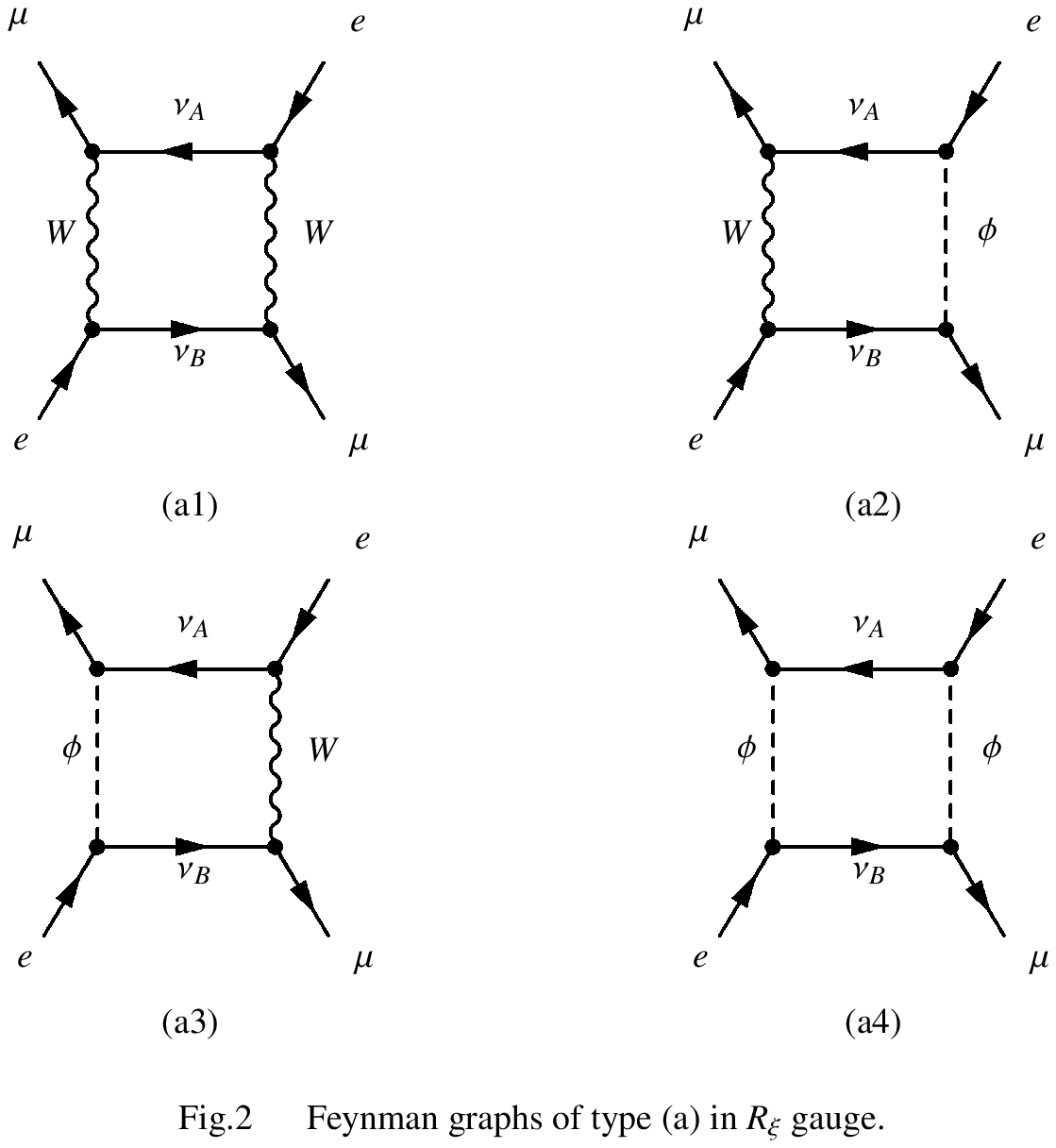}
\end{center}

A straightforward application of the $R_\xi$ gauge Feynman rules\cite{Tapei} to the above graphs yields the T-matrix elements
 \bea
T_{a1}=&&-\frac{g^4}{64\pi^2M^2_W}[\bar\mu(3)\gamma_\mu\frac{1-\gamma_5}{2}e(2)]
[\bar\mu(4)\gamma^\mu\frac{1-\gamma_5}{2}e(1)]\sum^6_{A=1}\sum^6_{B=1}(V_{\mu
A}V^\ast_{eA})(V_{\mu B}V^\ast_{eB})\cr &&\cdot\int^\infty_0 dt
\Bigg
[\frac{x_Ax_B}{(t+x_A)(t+x_B)(t+1)^2}\cdot\Big{\{}1+\frac{2(\xi-1)}{t+\xi}\cdot
t+\frac{(\xi-1)^2}{4(t+\xi)^2}\cdot t^2\Big{\}}\Bigg]\eea \bea
T_{a2}=T_{a3}=&&-\frac{g^4}{64\pi^2M^2_W}[\bar\mu(3)\gamma_\mu\frac{1-\gamma_5}{2}e(2)]
[\bar\mu(4)\gamma^\mu\frac{1-\gamma_5}{2}e(1)]\sum^6_{A=1}\sum^6_{B=1}(V_{\mu
A}V^\ast_{eA})(V_{\mu B}V^\ast_{eB})\cr &&\cdot\int^\infty_0 dt\Bigg
[\frac{x_Ax_B}{(t+x_A)(t+x_B)(t+1)(t+\xi)}\cdot\Big{\{}t+\frac{\xi-1}{4(t+\xi)}\cdot
t^2\Big{\}}\Bigg]\eea \bea
T_{a4}=&&-\frac{g^4}{64\pi^2M^2_W}[\bar\mu(3)\gamma_\mu\frac{1-\gamma_5}{2}e(2)]
[\bar\mu(4)\gamma^\mu\frac{1-\gamma_5}{2}e(1)]\sum^6_{A=1}\sum^6_{B=1}(V_{\mu
A}V^\ast_{eA})(V_{\mu B}V^\ast_{eB})\cr &&\cdot\int^\infty_0 dt\Bigg
[\frac{x_Ax_B}{(t+x_A)(t+x_B)(t+\xi)^2}\cdot\frac{t^2}{4}\Bigg]\eea
where $\bar\mu(3)=\bar\mu(p_3,s_3)$ , $\bar\mu(4)=\bar\mu(p_4,s_4)$
, $e(1)=e(p_1,s_1)$ and $e(2)=e(p_2,s_2)$ are the spinors of the muons
and electrons and $x_A=\frac{m^2_{\nu A}}{M^2_W}~, ~~~A=1,...,6$.
Note that in obtaining these results, we already  applied  properties (21) and
(22) to eliminate various self-cancelling terms. As such the integrals in (25)-(27) are finite even in the
$\xi\rightarrow\infty$ limit.

In order to discuss the $\xi$ dependence in a manifest way, we
rewrite these T-matrix elements as
\newline
\begin{center}
\includegraphics[width=14cm]{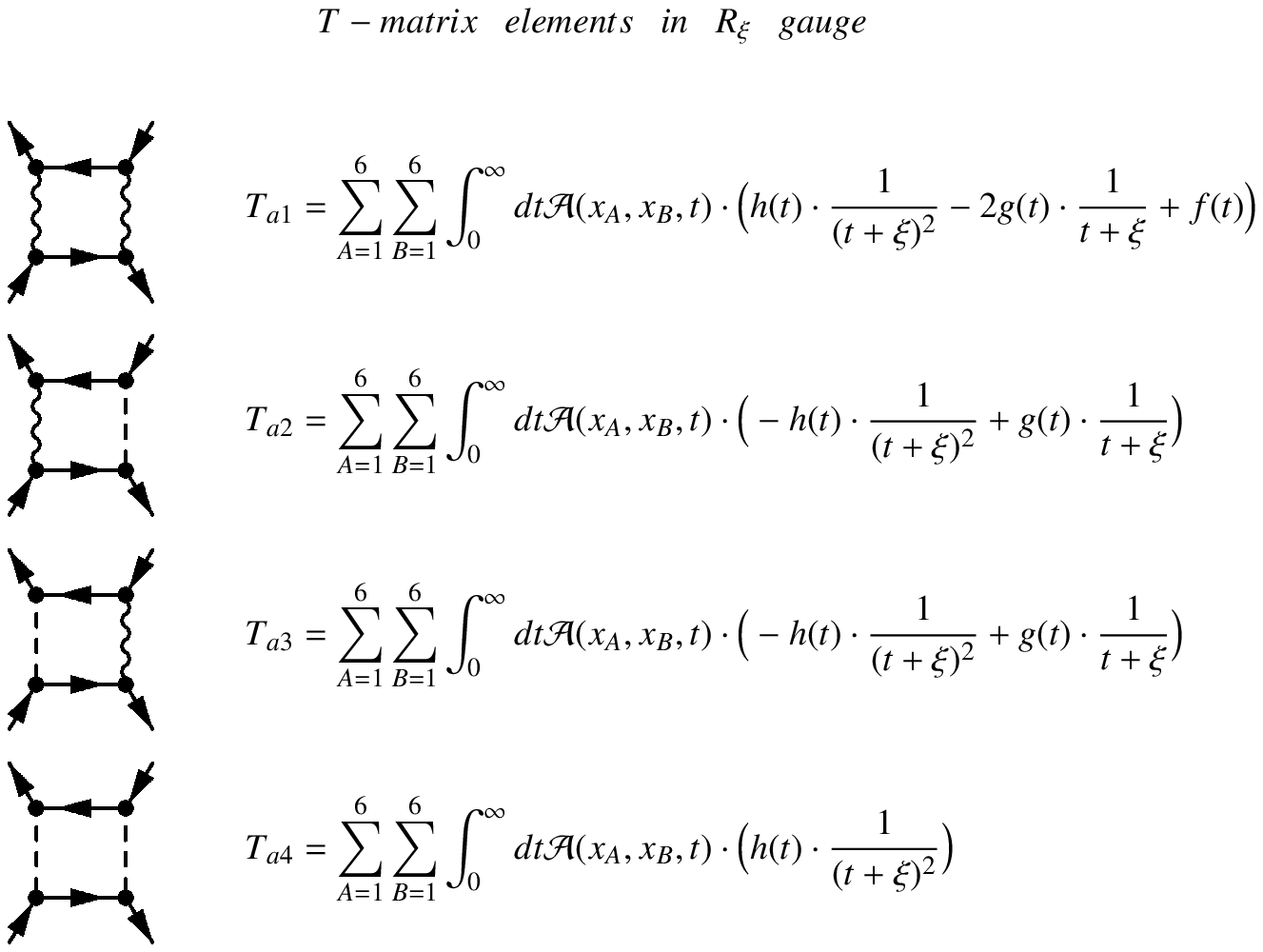}
\end{center}
where \bea \mathcal
{A}(x_A,x_B,t)=&&-\frac{g^4}{64\pi^2M^2_W}[\bar\mu(3)\gamma_\mu\frac{1-\gamma_5}{2}e(2)]
[\bar\mu(4)\gamma^\mu\frac{1-\gamma_5}{2}e(1)](V_{\mu
A}V^\ast_{eA})(V_{\mu B}V^\ast_{eB})\cr &&\cdot
\frac{x_Ax_B}{(t+x_A)(t+x_B)}\eea with \bea
h(t)=\frac{t^2}{4}~~~,~~~~
g(t)=\frac{t+\frac{t^2}{4}}{t+1}~~~and~~~~~
f(t)=\frac{1+2t+\frac{t^2}{4}}{(t+1)^2}\eea

Note that the $\frac{1}{(t+\xi)^2}$ terms from the second and
third graphs totally cancel against the ones from the first and
fourth graphs ,while the $\frac{1}{t+\xi}$ terms from the second and
third graphs exactly cancel the one from the first graph. All
$\xi$ dependent contributions thus vanish and the only remaining piece
is the term containing $f(t)$ from the first graph, which is $\xi$
independent. Hence, we have the gauge invariant T-matrix element for
graphs of type (a)  \bea
T_a=&&-\frac{g^4}{64\pi^2M^2_W}[\bar\mu(3)\gamma_\mu\frac{1-\gamma_5}{2}e(2)]
[\bar\mu(4)\gamma^\mu\frac{1-\gamma_5}{2}e(1)]\sum^6_{A=1}\sum^6_{B=1}(V_{\mu
A}V^\ast_{eA})(V_{\mu B}V^\ast_{eB})x_Ax_B\cr &&\cdot\int^\infty_0
dt\frac{1+2t+\frac{t^2}{4}}{(t+x_A)(t+x_B)(t+1)^2}\cr
=&&-\frac{G^2_FM^2_W}{8\pi^2}[\bar\mu(3)\gamma_\mu(1-\gamma_5)e(2)]
[\bar\mu(4)\gamma^\mu(1-\gamma_5)e(1)]\Bigg[\sum^6_{A=1}(V_{\mu
A}V^\ast_{eA})^2S(x_A)\cr &&+\sum^6_{A,B=1;A\neq B}(V_{\mu
A}V^\ast_{eA})(V_{\mu B}V^\ast_{eB})T(x_A,x_B)\Bigg]\eea Here we have introduced the Fermi scale \be
\frac{G_F}{\sqrt 2}=\frac{g^2}{8M^2_W}\ee along with the Inami-Lim\cite{Inami} function \be
S(x_A)=\frac{x^3-11x^2+4x}{4(1-x)^2}-\frac{3x^3}{2(1-x)^3}\ln (x)\ee
We have also defined \be
T(x_A,x_B)=x_Ax_B\Big(\frac{J(x_A)-J(x_B)}{x_A-x_B}\Big)=T(x_B,x_A)\ee
with \be
J(x)=\frac{x^2-8x+4}{4(1-x)^2}\ln{(x)}-\frac{3}{4}\frac{1}{(1-x)}\ee
\vspace{1cm}

In a similar manner, the graph (c) in Fig.1   represents  the following four
graphs:
\begin{center}
\includegraphics[width=11cm]{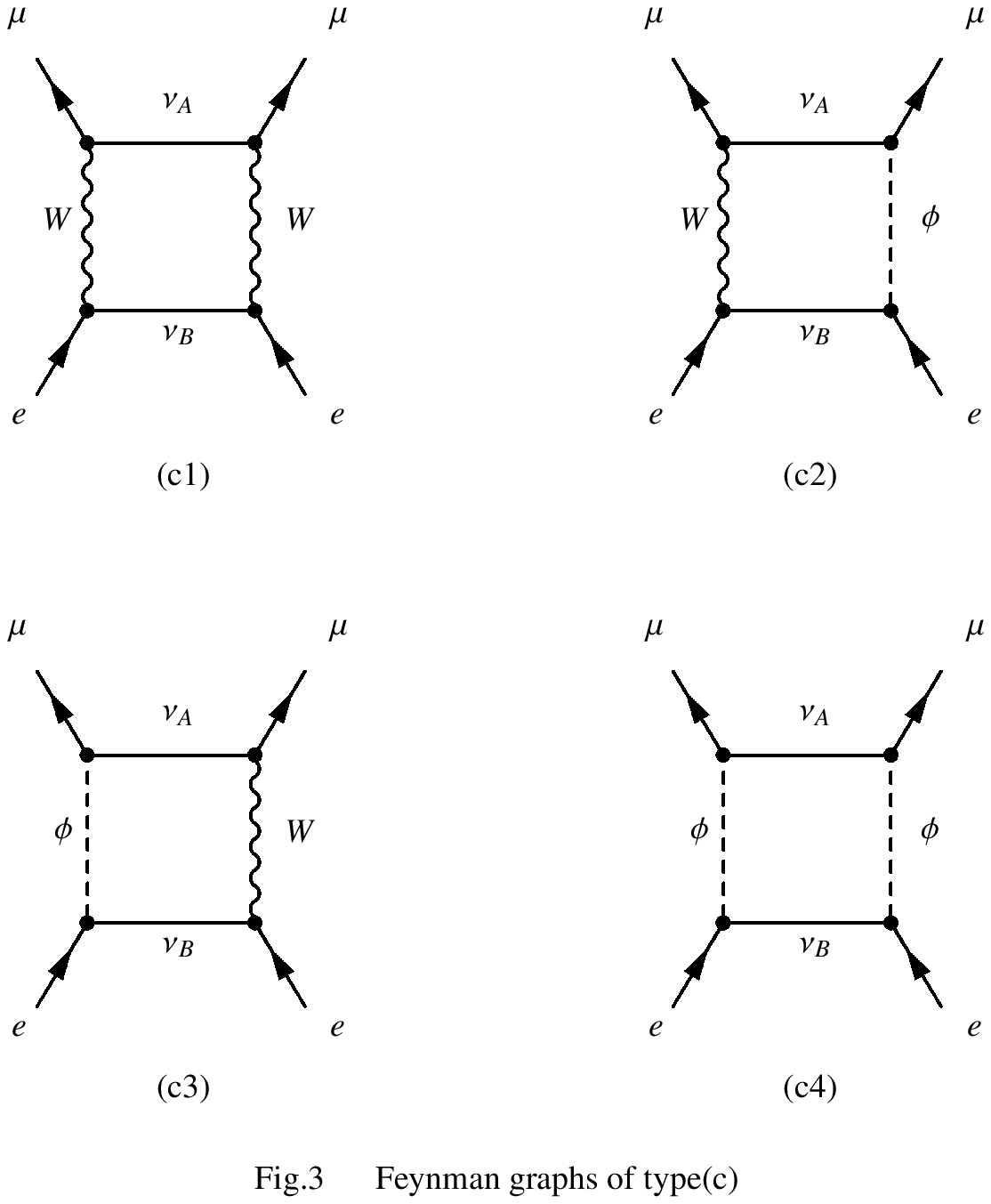}
\end{center}

The T-matrix elements of the above four graphs are
\begin{center}
\includegraphics[width=12cm]{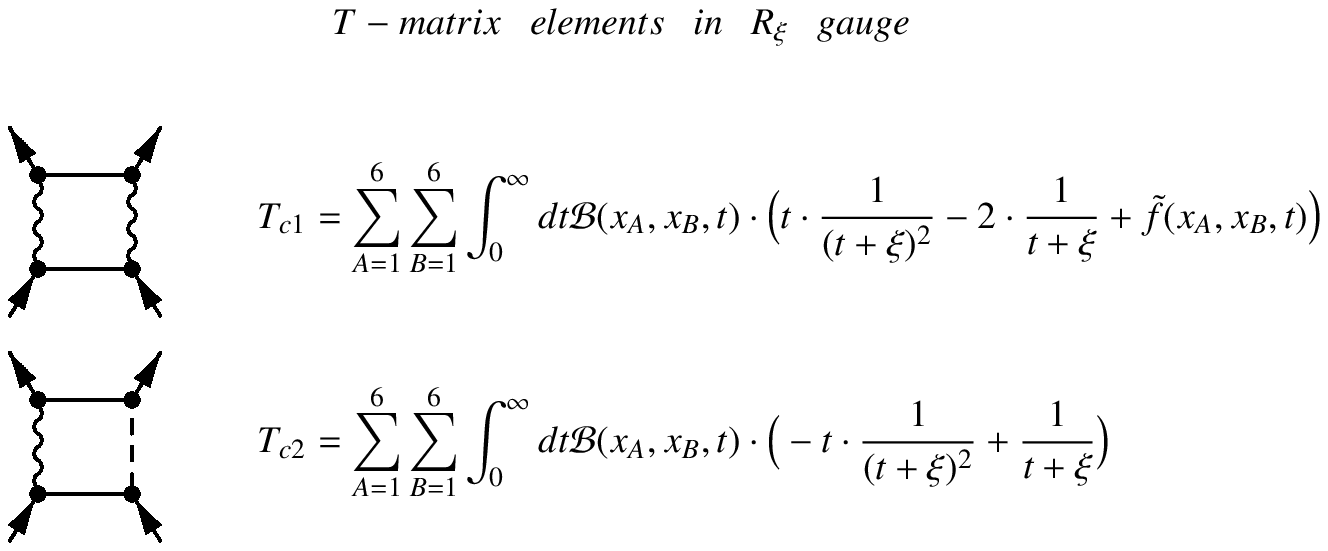}
\end{center}
\begin{center}
\includegraphics[width=11cm]{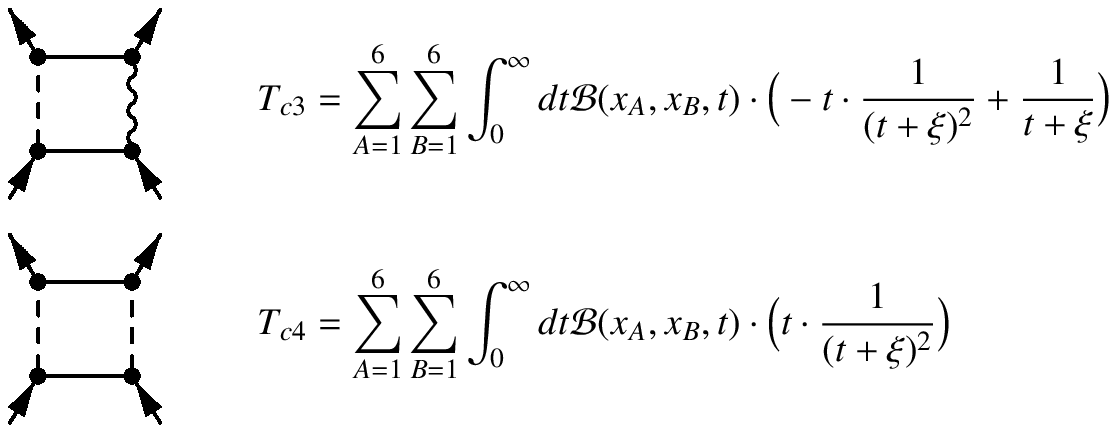}
\end{center}
where \bea \mathcal {B}(x_A,x_B,t)=&&\frac{g^4}{64\pi^2
M^4_W}[\bar\mu(3)\gamma^\mu \frac{1-\gamma_5}{2}e(2)]
[\bar\mu(4)\gamma_\mu\frac{1-\gamma_5}{2}e(1)](V_{\mu
A})^2(V_{eB}^\ast)^2\cr &&\cdot\frac{m_A\cdot m_B}{2}
\cdot\frac{x_Ax_B}{(t+x_A)(t+x_B)}\eea and\bea \tilde{f}
(x_A,x_B,t)=\frac{\frac{4t}{x_A\cdot x_B}+t+2}{(t+1)^2}\eea

In a similar fashion to the case for the graphs of type (a), all the $\xi$
dependent terms again cancel against each other leaving only the $\xi$ independent
 $\tilde{f} (x_A, x_B,t)$  term.
Thus the type (c) T-matrix element is gauge invariant and is given by \bea T_c=&&\frac{G^2_FM^2_W}{8\pi^2}[\bar\mu(3)\gamma^\mu
(1-\gamma_5)e(2)]
[\bar\mu(4)\gamma_\mu(1-\gamma_5)e(1)]\sum^6_{A=1}(V_{\mu
A})^2\sum^6_{B=1}(V_{eB}^\ast)^2\cr &&
\frac{\sqrt{x_Ax_B}}{2}\cdot\int^\infty_0
dt\Bigg{\{}\frac{4t+x_Ax_B(t+2)}{(t+x_A)(t+x_B)(t+1)^2}\Bigg{\}}\eea
If $x_A=x_B$, the relevant integral is \bea I(x_A)&=&\frac{\sqrt{x_Ax_B}}{2}\cdot\int^\infty_0
dt\frac{4t+x_Ax_B(t+2)}{(t+x_A)(t+x_B)(t+1)^2} \cr
&=&\frac{(x_A-4)x_A}{(x_A-1)^2}+\frac{(x_A^3-3x_A^2+4x_A+4)x_A}{2(x_A-1)^3}\ln{x_A}
\eea
while for $x_A\neq x_B$, it takes the form \bea
K(x_A,x_B)=&&\frac{\sqrt{x_Ax_B}}{2}\cdot\int^\infty_0
dt\frac{4t+x_Ax_B(t+2)}{(t+x_A)(t+x_B)(t+1)^2} \cr
=&&\sqrt{x_Ax_B}\frac{L(x_A,x_B)-L(x_B,x_A)}{x_A-x_B}\eea with\be
L(x_A,x_B)=\frac{4-x_Ax_B}{2(x_A-1)}+\frac{x_A(2x_B-x_Ax_B-4)}{2(x_A-1)^2}\ln
x_A\ee

The T-matrix element of graph (c) is thus secured as \bea
T_c=&&\frac{G^2_FM^2_W}{8\pi^2}[\bar\mu(3)\gamma^\mu
(1-\gamma_5)e(2)] [\bar\mu(4)\gamma_\mu(1-\gamma_5)e(1)]\cr
&&\cdot\Big[\sum^6_{A=1}(V_{\mu
A}V_{eA}^\ast)^2I(x_A)+\sum^6_{A,B=1;A\neq B}(V_{\mu
A})^2(V_{eB}^\ast)^2K(x_A,x_B)\Big]\eea

\vspace{0.5cm}

Combining the various contributions, the  T-matrix element can be reproduced using the gauge invariant
effective Lagrangian given by: \be {\cal L}_{eff}=\frac{G_{\bar MM}}{\sqrt
2}[\bar\mu\gamma^\mu (1-\gamma_5)e]
[\bar\mu\gamma_\mu(1-\gamma_5)e]\label{Leff}\ee where \bea \frac{G_{\bar
MM}}{\sqrt 2}=&&-\frac{G^2_FM^2_W}{16\pi^2}\Bigg[\sum^6_{A=1}(V_{\mu
A}V^\ast_{eA})^2S(x_A)+\sum^6_{A,B=1;A\neq B}(V_{\mu
A}V^\ast_{eA})(V_{\mu B}V^\ast_{eB})T(x_A,x_B)\cr
&&-\sum^6_{A=1}(V_{\mu A}V_{eA}^\ast)^2I(x_A)-\sum^6_{A,B=1;A\neq
B}(V_{\mu A})^2(V_{eB}^\ast)^2K(x_A,x_B)\Bigg]\cr
=&&-\frac{G^2_FM^2_W}{16\pi^2}\Bigg[\sum^6_{A=1}(V_{\mu
A}V^\ast_{eA})^2\Big(S(x_A)-I(x_A)\Big)\cr &&+\sum^6_{A,B=1;A\neq
B}\Big((V_{\mu A}V^\ast_{eA})(V_{\mu
B}V^\ast_{eB})T(x_A,x_B)-(V_{\mu
A})^2(V_{eB}^\ast)^2K(x_A,x_B)\Big)\Bigg]\label{GMM}\eea

\section{Limit on $M_R$}
Muonium (antimuonium) is a nonrelativistic Coulombic bound state of
an electron and an anti-muon (positron and muon). The nontrivial
mixing between the muonium ( $|M >$ ) and antimuonium ($|\bar M >$)
states is encapsulated in the effective Lagrangian of Eq. (\ref{Leff}) and
leads to the mass diagonal states given by the linear combinations
\be
|M_\pm>=\frac{1}{\sqrt{2(1+|\varepsilon|^2)}}[(1+\varepsilon)|M>\pm(1-\varepsilon)|\bar
M>]\ee where \be \varepsilon=\frac{\sqrt{\mathcal M_{M\bar
M}}-\sqrt{\mathcal M_{\bar MM}}}{\sqrt{\mathcal M_{M\bar
M}}+\sqrt{\mathcal M_{\bar MM}}} \ee \be \mathcal M_{M\bar
M}=\frac{<M|-\int d^3r\mathcal L_{eff}|\bar M>}{\sqrt{<M|M><\bar
M|\bar M>}},~~~\mathcal M_{\bar MM}=\frac{<\bar M|-\int d^3r\mathcal
L_{eff}|M>}{\sqrt{<M|M><\bar M|\bar M>}}\ee

Since the neutrino sector is expected to be CP violating, these will
be independent, complex matrix elements. If the neutrino sector
conserves CP, with $|M >$ and $|\bar M >$ CP conjugate states, then
$\mathcal M_{M\bar M}=\mathcal M_{\bar MM}$ and $\epsilon= 0$. In
general, the magnitude of the mass splitting between the two mass
eigenstates is
 \bea |\Delta M|=2\left|Re\sqrt{\mathcal M_{M\bar M}\mathcal M_{\bar MM}}~~\right|
\eea Since muonium and antimuonium are linear combinations of the
mass diagonal states, an initially prepared muonium or antimuonium
state will undergo oscillations into one another as a function of
time. The muonium-antimuonium oscillation time scale, $\tau_{\bar
MM}$, is given by \be\frac{1}{\tau_{\bar MM}}=|\Delta M|.\ee

We would like to evaluate $|\Delta M|$ in the nonrelativistic limit.
A nonrelativistic reduction of the effective Lagrangian of Eq.
(\ref{Leff}) produces the local, complex effective potential \bea
V_{eff}(\textbf{r})=8\frac{G_{\bar MM}}{\sqrt
2}\delta^3(\textbf{r})\eea

Taking the muonium (anitmuonium) to be in their respective Coulombic
ground states, $\phi_{100}(\textbf{r})= \frac{1}{\sqrt{\pi a^3_{\bar
MM}}}e^{-r/a_{\bar MM}}$, where $a_{\bar
MM}=\frac{1}{m_{red}\alpha}$ is the muonium Bohr radius with
$m_{red}=\frac{m_em_\mu}{m_e+m_\mu}\simeq m_e$ the reduced mass of
muonium, it follows that \bea \frac{1}{\tau_{\bar MM}}\simeq &&2\int
d^3r\phi^\ast_{100}(\textbf{r})|ReV_{eff}(\textbf r)|\phi(\textbf
r)_{100}\cr=&& 16\frac{|ReG_{\bar MM}|}{\sqrt
2}|\phi_{100}(0)|^2=\frac{16}{\pi}\frac{|ReG_{\bar MM}|}{\sqrt
2}\frac{1}{a^3_{\bar MM}}\eea Thus we secure an oscillation time
scale\be \frac{1}{\tau_{\bar MM}}\simeq
\frac{16}{\pi}\frac{|ReG_{\bar MM}|}{\sqrt 2} m_e^3\alpha^3\ee

The present experimental limit\cite{Willmann} on the non-observation
of muonium-antimuonium oscillation translates into the bound
$|ReG_{\bar MM}|\leq 3.0\times 10^{-3}G_F$ where
$G_F\simeq1.16\times10^{-5}GeV^{-2}$ is the Fermi scale. This limit
can then be used to construct a crude lower bound on $M_R$. For the
case when the neutrino masses arise from a see-saw mechanism and
taking $m_D$ to be of order $M_W$, the $M_R$ dependence of $G_{\bar
MM}$ is obtained from Eq. (\ref{GMM}) as: \bea &&Case
1:~~~~|ReG_{\bar MM}|\sim
\frac{G^2_FM^4_W}{M^2_R}\ln{\frac{M_R}{M_W}},
~~~~~~~A=1,2,3,~~B=1,2,3\cr &&Case 2:~~~~|ReG_{\bar MM}|\sim
\frac{G^2_FM^4_W}{M^2_R}\ln{\frac{M_R}{M_W}},
~~~~~~~A=4,5,6,~~B=4,5,6\cr &&Case 3:~~~~|ReG_{\bar MM}|\sim
\frac{G^2_FM^6_W}{M^4_R}\ln{\frac{M_R}{M_W}},
~~~~~~~A=1,2,3,~~B=4,5,6 \eea

Case 1 and case 2 give the same order $M_R$
dependence, while case 3 is
suppressed by an additional factor of $\frac{M^2_W}{M^2_R}$. Hence, the term
 $\frac{G^2_FM^4_W}{M^2_R}\ln{\frac{M_R}{M_W}}$ gives
the dominant contribution. We then roughly calculate a bound of
$M_R$ as \bea \frac{G^2_FM^4_W}{M^2_R}\ln{\frac{M_R}{M_W}}\leq
3.0\times 10^{-3}G_F\eea which has also been obtained in reference
\cite{Kim}. Using $M_W\simeq 80.4$ GeV and
$G_F=1.166\times10^{-5}GeV^{-2}$, we finally secure \be
M_R\geq6\times10^2GeV\ee Note that this is just a rough estimate
since we are retaining only the  dependence on $M_R$ while
neglecting all numerical dependence on the mixing angles and CP
violating phases in $V_{aA}$.
\begin{center}\section*{Acknowledgements}\end{center} It's a pleasure to thank Professor
S. T. Love, Professor T. E. Clark and Chi Xiong for useful
discussions.
\newpage
\section*{Appendix: Proof of Identity (22)}
Using the definition of the mixing matrix
$V_{aA}=\sum^3_{c=1}(A^{-1}_L)_{ac}U_{cA}$,
one can write \bea
\sum^6_{A=1}V_{aA}V_{bA}m_{\nu A}&=&\sum^6_{A=1}\left(\sum^3_{c=1}(A^{-1}_L)_{ac}U_{cA}\right)\cdot\left(\sum^3_{d=1}(A^{-1}_L)_{bd}U_{dA}\right)\cdot
m_{\nu
A}\cr&=&\sum^3_{c=1}\sum^3_{d=1}(A^{-1}_L)_{ac}\left(\sum^6_{A=1}U_{cA}m_{\nu
A}U_{dA}\right)(A^{-1}_L)_{bd}\eea where $m_{\nu A}$ are the
diagonal elements of matrix $M^\nu_{diag}$, \bea m_{\nu
A}=(M^\nu_{diag})_{AA}.\eea Consequently, we can express
$\sum^6_{A=1}U_{cA}m_{\nu A}U_{dA}$ as a product of matrices and
equation (53) takes the form \bea\sum^6_{A=1}V_{aA}V_{bA}m_{\nu
A}=\sum^3_{c=1}\sum^3_{d=1}(A^{-1}_L)_{ac}\left(U M^\nu_{diag
}U^T\right)_{cd}(A^{-1}_L)_{bd}\eea Using equation (9), $
M^\nu_{diag}=U^T M^\nu U$, it follows that
\bea M^{\nu\ast}=U M^\nu_{diag} U^T\eea Substituting this result
back into eq. (55) then gives \be \sum^6_{A=1}V_{aA}V_{bA}m_{\nu A}
=\sum^3_{c=1}\sum^3_{d=1}(A^{-1}_L)_{ac}(M^{\nu\ast})_{cd}(A^{-1}_L)_{bd}\ee
where \be M^{\nu\ast}=\left(\begin{array}{c}0~~~~~~
(m^D)^{T\ast}\\m^{D\ast}~~~~~~m^{R\ast}\end{array}\right)\ee Since c
and d both run from 1 to 3, ~~$(M^{\nu\ast})_{cd}$ are the elements
of the upper left $3\times3$ block of matrix (58), which is zero.
Hence, we secure the identity \be\sum^6_{A=1}V_{aA}V_{bA}m_{\nu
A}=0\ee

\end{document}